\documentclass{article}

\usepackage{amsmath}
\usepackage{amssymb}
\usepackage[dvips]{geometry}
\usepackage{fullpage}
\usepackage[T1]{fontenc}
\usepackage{graphicx}
\usepackage{url}
\usepackage{citesort}
\usepackage[amssymb]{SIunits}

\newcommand{\J}{\mathrm{j}}  
\newcommand{\D}{\mathrm{d}}  
\newcommand{\E}{\mathrm{e}}  

\newcommand{\hankel}[1]{H_{#1}^{(1)}}
\newcommand{\besselj}[1]{J_{#1}}
\newcommand{\lfn}[1]{\mathcal{L}_{#1}}

\begin{document}

\title{The radiating part of circular sources}
\author{Michael Carley\\ Department of Mechanical
  Engineering, University of Bath, Bath BA2 7AY, England\\
  m.j.carley@bath.ac.uk} 

\date{\today}

\maketitle

\begin{abstract}
  An analysis is developed linking the form of the sound field from a
  circular source to the radial structure of the source, without
  recourse to far-field or other approximations. It is found that the
  information radiated into the field is limited, with the limit fixed
  by the wavenumber of source multiplied by the source radius
  (Helmholtz number). The acoustic field is found in terms of the
  elementary fields generated by a set of line sources whose form is
  given by Chebyshev polynomials of the second kind, and whose
  amplitude is found to be given by weighted integrals of the radial
  source term. The analysis is developed for tonal sources, such as
  rotors, and, for Helmholtz number less than two, for random disk
  sources. In this case, the analysis yields the cross-spectrum
  between two points in the acoustic field. The analysis is applied to
  the problems of tonal radiation, random source radiation as a model
  problem for jet noise, and to noise cancellation, as in active
  control of noise from rotors. It is found that the approach gives an
  accurate model for the radiation problem and explicitly identifies
  those parts of a source which radiate. 
\end{abstract}


\section{Introduction}
\label{sec:intro}

A problem in many applications is that of sound generated by circular
sources. These include rotors of various types such as aircraft
propellers and fans, wind turbines and cooling fans; vibrating systems
such as loudspeakers; ducts such as aircraft engines, ventilation
systems and exhausts; and distributed sources with axial symmetry such
as jets. There have been numerous studies of the noise generation and
radiation process in each of these areas extending over many
decades. These studies can be divided into those which examine the
relationship between the acoustic source and the physical processes
which give rise to it, for example the work of
Lighthill~\cite{lighthill52} and of Ffowcs Williams and
Hawkings~\cite{ffowcs-williams-hawkings69b} which relate aerodynamic
quantities to acoustic sources, and those which examine the radiated
field generated by a given source distribution, such as methods for
prediction of the field radiated by pistons and
loudspeakers~\cite{oberhettinger61a,pierce89,mellow06,mellow08} or from
a known rotating source
distribution~\cite{gutin48,wright69,chapman93,carley99}.

There are a number of areas where these issues, those of generation
and radiation, overlap. One is the general area of source
identification. There have been many attempts to develop methods which
use acoustic measurements to infer, in greater or lesser detail, the
source distribution responsible for the acoustic field. In the case of
rotating sources, some examples include cooling
fans~\cite{gerard-berry-masson05a,gerard-berry-masson05b,%
  gerard-berry-masson-gervais07} and
propellers~\cite{peake-boyd93,li-zhou95,%
  minniti-blake-mueller01a,minniti-blake-mueller01b}, while a number
of groups have developed methods for the inverse problem for radiation
from a duct termination~\cite{holste-neise97,lewy05,lewy08,%
  castres-joseph07a,castres-joseph07b}. Such studies can have a number
of motivations. The first is to use near-field data, for example those
taken in wind-tunnel tests, to predict the far acoustic field. In this
case, the requirement is to extract information about source strength
and directivity, but there is no need to know which processes generate
the source. A second motivation, however, is the identification of the
noisiest parts of the source with a view to reduction of noise at
source, for example the identification of ``hot spots'' caused by
unsteady loading on a cooling
fan~\cite{gerard-berry-masson05a,gerard-berry-masson05b}. In this case,
the link between the aerodynamics and the source is an essential part
of the solution of the problem.

In each of the applications of source identification listed, the
authors have recognized that the problem is (very)
ill-conditioned. This can be attributed to physical causes, and is not
merely an artifact of the methods used. Recent
analysis~\cite{carley09,carley10b,carley10c} has given a framework for
the study of this ill-conditioning by quantifying the source
information which is radiated into the acoustic near and far
fields. As described below, it has been found that the source can be
decomposed into orthogonal modes based on Chebyshev polynomials, only
a limited number of which radiate a detectable acoustic field, with
the limit being fixed by the source frequency. 

A second area where the issues of generation and radiation overlap is
that of jet noise. Lighthill's acoustic analogy~\cite{lighthill52} is
accepted as an exact theory for noise generation by turbulence and
there is solid evidence for the validity of his source term, as
demonstrated by high quality numerical simulation~\cite{freund01}. This
knowledge, however, is not sufficient to explain certain features of
jet noise, in particular the low radiation efficiency of subsonic jets
and the low order structure of the acoustic field. It is known that
subsonic jets radiate only a small fraction of the source energy, a
view given support by the very small changes in the flow which suffice
to give large reductions in noise, when control is
applied~\cite{freund10}. It is also known that the acoustic far field
of a jet is significantly simpler than the flow field. In a recent
study~\cite{jordan-schlegel-stalnov-noack-tinney07}, modal
decomposition of the far-field noise and of the flow field of a
Mach~0.9 jet showed that~24 modes were sufficient to capture~90\% of
the energy of the acoustic field, but~350 were required to
resolve~50\% of the flow energy. Clearly, a very large part of the
flow, however energetic it might be, simply does not radiate but it is
not obvious if this is due to the nature of the source or purely a
result of radiation effects.

The radiation effect has been explained in terms of source
cancellation~\cite{michel07,michel09} and by viewing the radiation
process as equivalent to the imposition of a spatial filter using a
wavenumber criterion. Such an approach has been used by
Freund~\cite{freund01} who found that the part of the source which
radiates is indeed that part left over after applying an appropriate
spatial filter. Similarly, Sinayoko and
Agarwal~\cite{sinayoko-agarwal10} apply a linear convolution filter to
decompose the flow into radiating and non-radiating parts.

The analysis to be presented below attempts to explain some of these
features. Previous work~\cite{carley09,carley10b,carley10c} has found
limits on the information radiated from a tonal circular source,
motivated by a desire to understand the ill-conditioning of source
identification methods. These limits have been found without recourse
to a far-field approximation, making the approach suitable for
analysis of general problems. The remainder of this paper contains an
extension of the theory to explicitly include the radial source term,
and to yield spectral quantities in the acoustic field of random
sources. 

The first extension, which can be viewed as a generalization of
previous work on axisymmetric radiators~\cite{stepanishen76}, will help
explain radial cancellation effects, which have been studied in jet
noise using a far-field formulation~\cite{michel09} but not, to the
author's knowledge, in the near field. It will be found that for a
given azimuthal order, many different sources radiate identical
acoustic fields, differing only by a scaling factor. This result is
part of the explanation for the ill-conditioning of identification
methods and also opens a possible approach to the development of
control systems by identifying a class of sources which can give rise
to practically identical acoustic fields.

The second extension, to predicting the cross-spectrum between the
acoustic pressures radiated by a random source to arbitrary points in
the near and/or far field, is an extension of an earlier ring-source
model for radiation from random sources characteristic of
jets~\cite{michalke83}. In this case, it will be found that the
cross-spectrum depends on four constants, functions of observer
radial separation, which are weighted integrals of the source
cross-spectrum. 

The results to be presented arise from two different exact theories
for radiation from circular
sources~\cite{carley10,carley10b,carley10c} which are combined to give
a formulation for the information in the acoustic field in terms of
radiation functions and weighted integrals of the source term. The
implications of the results are discussed in terms of the information
content of the acoustic field and with regard to some of the
measurement methods used to study noise sources.

\section{Tonal disk source}
\label{sec:analysis}

\begin{figure}
  \centering
  \includegraphics{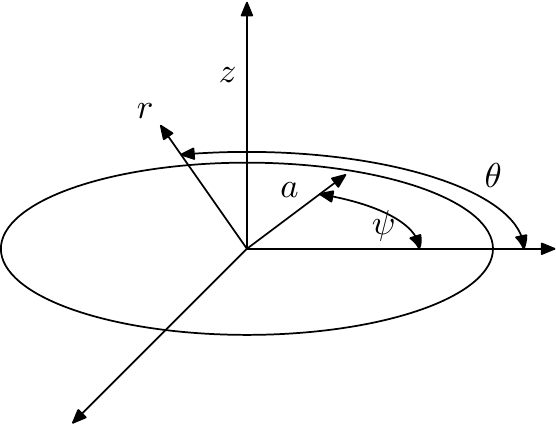}
  \caption{Coordinate system for disk radiation calculations}
  \label{fig:coordinate}
\end{figure}

The problem is initially formulated as that of calculating the
acoustic field radiated by a monopole source distributed over a
circular disk. The system for the analysis is shown in
Figure~\ref{fig:coordinate} with cylindrical coordinates
$(r,\theta,z)$ for the observer and $(a,\psi,0)$ for the source. All
lengths are non-dimensionalized on disk radius. The field from one
azimuthal mode of the acoustic source, specified as $s_{n}(a)\exp\J [n
\psi-\omega t]$, has the form $P_{n}(k,r,z)\exp\J [n \theta-\omega
t]$, with $P_{n}$ given by the Rayleigh
integral~\cite{goldstein74,carley09}:
\begin{align}
  \label{equ:disk}
  P_{n}(k,r,z) &= 
  \int_{0}^{1}
  \int_{0}^{2\pi} \frac{\E^{\J(kR'+n\psi)}}{4\pi
    R'}\,\D  \psi s_{n}(a) a\,\D a,\\
  R' &= 
  \left[
    r^{2} + a^{2} - 2ra\cos\psi + z^{2}
  \right]^{1/2},\nonumber
\end{align}
where $k$ is non-dimensional wavenumber (Helmholtz number).  


\subsection{Equivalent line source expansion}
\label{sec:line}

\begin{figure}
  \centering
  \includegraphics{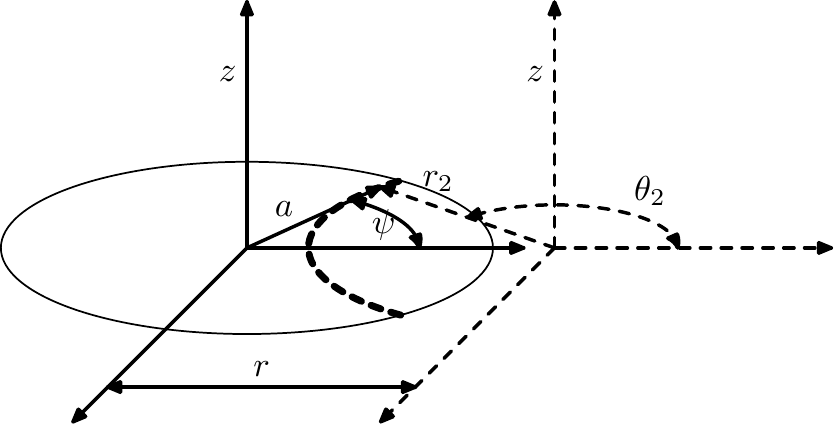}
  \caption{Transformation to equivalent line source}
  \label{fig:sideline}
\end{figure}

The analysis of the nature of the sound field from an arbitrary disk
source is based on a transformation of the disk to an exactly
equivalent line source, an approach which has been used to study
transient radiation from pistons~\cite{oberhettinger61a,pierce89},
rotor noise~\cite{chapman93,carley99} and source identification
methods~\cite{carley09,carley10c,carley10b}.

The transformation to a line source is shown in
Figure~\ref{fig:sideline}, which shows the new coordinate system
$(r_{2},\theta_{2},z)$ centred on a sideline of constant radius
$r$. Under this transformation:
\begin{align}
  \label{equ:transformed}
  P_{n}(k,r,z) &= \int_{r-1}^{r+1} \frac{\E^{\J kR'}}{R'}
  K(r,r_{2})r_{2}\,\D
  r_{2},\\
  R' &= \left(r_{2}^{2} + z^{2}\right)^{1/2},\nonumber\\
  \label{equ:kfunc}
  K(r,r_{2}) &= \frac{1}{4\pi}
  \int_{\theta_{2}^{(0)}}^{2\pi-\theta_{2}^{(0)}} \E^{\J
    n\psi}s_{n}(a)\,\D\theta_{2},
\end{align}
for observer positions with $r>1$, with the limits of integration
given by:
\begin{align}
  \label{equ:theta}
  \theta_{2}^{(0)} &= \cos^{-1}\frac{1-r^{2}-r_{2}^{2}}{2rr_{2}}.
\end{align}
Functions of the form of $K(r,r_{2})$ have been analyzed in previous
work~\cite{carley99} and can be written:
\begin{align}
  \label{equ:kfunc:exp}
  K(r,r_{2}) &= \sum_{q=0}^{\infty} u_{q}(r)U_{q}(s)(1-s^2)^{1/2},
\end{align}
where $U_{q}(s)$ is a Chebyshev polynomial of the second kind,
$s=r_{2}-r$ and the coefficients $u_{q}(r)$ are functions of $r$ but
not of $z$. Inserting Equation~\ref{equ:kfunc:exp} into
Equation~\ref{equ:transformed}:
\begin{align}
  \label{equ:line:ip}
  P_{n}(k,r,z) &=
  \sum_{q=0}^{\infty} u_{q}(r)\lfn{q}(k,r,z),\\
  \label{equ:lfunc}
  \lfn{q}(k,r,z) &= \int_{-1}^{1} \frac{\E^{\J kR'}}{R'}
  U_{q}(s) (r+s)(1-s^2)^{1/2}\,\D s,\\
  R' &= \left[(r+s)^{2} + z^{2}\right]^{1/2}.
\end{align}
The radiation properties of the integral of Equation~\ref{equ:lfunc}
have been examined in some detail elsewhere~\cite{carley10b,carley10c},
giving an exact result for the in-plane case $z=0$:
\begin{align}
  \label{equ:line:series}
  \lfn{q}(k,r,0) &= 
  \J^{q} (q+1) \pi \E^{\J k r}  \frac{\besselj{q+1}(k)}{k}.
\end{align}
For large order $q$, the Bessel function $J_{q}(k)$ is exponentially
small for $k<q$ so that the line source modes with order $q>k$
generate noise fields of exponentially small amplitude. Since the
integrals have their maximum in the plane $z=0$,
Equation~\ref{equ:line:series} says that the whole field is of
exponentially small amplitude. This gives an indication of how much of
a given source distribution radiates into the acoustic field, near or
far.

In previous analyses, two approximations to $\lfn{q}$ have been
developed. One is an asymptotic formula valid in the limit
$k\to\infty$, derived using the method of stationary
phase~\cite{carley10b,carley10c}. This will not be required here, but
we will make use of the far-field form of Equation~\ref{equ:lfunc}:
\begin{align}
  \label{equ:lfunc:ff}
  \lfn{q} &\approx \J^{q}\pi\frac{\E^{\J k R}}{R}\frac{q+1}{k\sin\phi}
  \biggl[
  \left(r+\J\frac{q+2}{k\sin\phi}\right)
  J_{q+1}(k\sin\phi)
  -\J J_{q}(k\sin\phi)
  \biggr],
\end{align}
where $R=[r^{2}+z^{2}]^{1/2}$ and $\phi=\cos^{-1}z/R$.

Given the basic information about the form of the radiated field,
there remains to establish the relationship between the radial
structure of the source $s_{n}(a)$ and the line source coefficients
$u_{q}(r)$.

\subsection{Series expansion for spinning sound fields}
\label{sec:series}

A recently derived series~\cite{carley10} for the field radiated by a
ring source of radius $a$ can be used to find a second expression for
the sound radiated by a disk source with arbitrary radial variation:
\begin{align}
  R_{n} &= \int_{0}^{2\pi} \frac{\E^{\J(kR'+n\psi)}}{4\pi R'}\,\D\psi,
  \nonumber\\
  &=
  \J^{2n+1}\frac{\pi}{4}
  \frac{1}{(aR)^{1/2}}
  \sum_{m=0}^{\infty}
  (-1)^m
  \frac{(2n+4m+1)(2m-1)!!}{(2n+2m)!!}\nonumber\\
  \label{equ:ring}
  &\times
  \hankel{n+2m+1/2}(kR)  P_{n+2m}^{n}(\cos\phi)
  \besselj{n+2m+1/2}(ka),
\end{align}
with $\hankel{\nu}(x)$ the Hankel function of the first kind of order
$\nu$, $\besselj{\nu}$ the Bessel function of the first kind and
$P_{n}^{m}$ the associated Legendre function. The observer position is
specified in the spherical polar coordinates used in
Equation~\ref{equ:lfunc:ff}.

Multiplication by the radial source term $a s_{n}(a)$ and
integration gives an expression for the field radiated by a general
source of unit radius and azimuthal order $n$:
\begin{align*}
  P_{n}(k,r,z) &=
  \J^{2n+1}\frac{\pi}{4}
  \sum_{m=0}^{\infty}
  (-1)^m
  \frac{(2n+4m+1)(2m-1)!!}{(2n+2m)!!} P_{n+2m}^{n}(\cos\phi)S_{n+2m},\\
  S_{n+2m}(k,r,z) &= \int_{0}^{1} s_{n}(a)
  \besselj{n+2m+1/2}(ka)\hankel{n+2m+1/2}(kR)  
  \left(
    \frac{a}{R}
  \right)^{1/2}\,\D a.
\end{align*}

Setting $z=0$ ($\phi=\pi/2$, $R=r$):
\begin{align}
  \label{equ:series:ip}
  P_{n}(k,r,0) &=
  \frac{\J\pi}{4}
  \sum_{m=0}^{\infty}
  A_{m}S_{n+2m},\\
  A_{m} &= 
  \frac{1}{m!}
  \frac{(2n+4m+1)(2n+2m-1)!!(2m-1)!!}{2^{m} (2n+2m)!!},\nonumber
\end{align}
where use has been made of the expression~~\cite{gradshteyn-ryzhik80}:
\begin{align}
  P_{n+2m}^{n}(0) = \frac{(-1)^{m+n}}{2^{m}}\frac{(2n+2m-1)!!}{m!}.
\end{align}

\subsection{Line source coefficients}
\label{sec:coefficients}

The expressions for $P_{n}$ from section~\ref{sec:line} and
section~\ref{sec:series} are both exact and can be equated to derive a
system of equations relating the coefficients $u_{q}(r)$ to the
weighted integrals of the radial source distribution $s_{n}(a)$:
\begin{align}
  \frac{\J}{4}
  \sum_{m=0}^{\infty}
  A_{m}
  S_{n+2m}
  &=
  \label{equ:system}
  \E^{\J k r}
  \sum_{q=0}^{\infty} u_{q}(r)
  \J^{q} (q+1) \frac{\besselj{q+1}(k)}{k}.
\end{align}
Under repeated differentiation, Equation~\ref{equ:system} becomes a
lower triangular system of linear equations which connects the
coefficients $u_{q}(r)$ and $S_{n+2m}$:
\begin{align}
  \frac{\J}{4}
  \sum_{m=0}^{\infty}
  A_{m}
  S_{n+2m}^{(v)} &=
  \label{equ:diff}
  \sum_{q=0}^{\infty} u_{q}(r)
  \J^{q} (q+1)
  \left[
    \E^{\J k r}
    \frac{\besselj{q+1}(k)}{k}
  \right]^{(v)},
\end{align}
where superscript $(v)$ denotes the $v$th partial derivative with
respect to $k$, evaluated at $k=0$.

Using standard series~\cite{gradshteyn-ryzhik80}, the products of
special functions can be written:
\begin{align}
  \label{equ:prod:exp:j}
    \E^{\J k r}\frac{\besselj{q+1}(k)}{k} &=
    \frac{1}{\J^{q}} \sum_{t=0}^{\infty}(\J k)^{t+q} E_{t,q}(r),\\
    E_{t,q}(r) &=  \frac{1}{2^{q+1}}
    \sum_{s=0}^{[t/2]}
    \frac{r^{t-2s}}{4^{s}s!(s+q+1)!(t-2s)!},\nonumber
\end{align}
where $[t/2]$ is the largest integer less than or equal to $t/2$, and
\begin{align}
  \left(
    \frac{a}{r}
  \right)^{1/2}
  \hankel{n+1/2}(kr)\besselj{n+1/2}(ka) &= 
  \left(
    \frac{r}{2}
  \right)^{2n+1}
  \sum_{t=0}^{\infty}
  \frac{k^{2t+2n+1}}{t!}
  \left(
    -\frac{r^{2}}{4}
  \right)^{t}V_{n,t}(a/r) \nonumber\\
  &- (-1)^n\J \sum_{t=0}^{\infty}
  \frac{k^{2t}}{t!}
  \left(
    -\frac{r^{2}}{4}
  \right)^{t}W_{n,t}(a/r),
  \label{equ:prod:h:j}
\end{align}
with the polynomials $V_{n,t}$ and $W_{n,t}$ given by:
\begin{subequations}
  \label{equ:vwpoly}
  \begin{align}
    V_{n,t}(x) &= \sum_{s=0}^{t} {t \choose s}
    \frac{x^{2s+n+1}}{\Gamma(n+s+3/2)\Gamma(t-s+n+3/2)},\\
    W_{n,t}(x) &= \sum_{s=0}^{t} {t \choose s}
    \frac{x^{2s+n+1}}{\Gamma(n+s+3/2)\Gamma(t-s-n+1/2)}.
  \end{align}
\end{subequations}

Given the power series, the derivatives at $k=0$ are readily found:
\begin{subequations}
  \label{equ:derivatives}
  \begin{align}
    \J^{q}
    \frac{\partial^{v}}{\partial k^{v}}
    \left[
      \E^{\J k r}\frac{\besselj{q+1}(k)}{k}
    \right]_{k=0}
    &= 
    \left\{
    \begin{array}{ll}
      0, & v < q;\\
      \J^{v}v!E_{v-q,q}(r), & v \geq q.
    \end{array}
    \right.\\
      \frac{\partial^{v}}{\partial k^{v}}
     \left[
      (a/r)^{1/2}
       \hankel{n+1/2}(kr)\besselj{n+1/2}(ka)
     \right]_{k=0}
     &= \nonumber\\
     \left\{
     \begin{array}{lll}
       \displaystyle
       0, & v=2v'+1,& v' < n;\\
       \displaystyle
       \left(
         \frac{r}{2}
       \right)^{2n+1}
       \left(
         -\frac{r^{2}}{4}
       \right)^{v'-n}
       \frac{v!}{(v'-n)!}V_{n,v'-n}(a/r), & v=2v'+1, & v'\geq n;\\
       \displaystyle
       -(-1)^{n}\J \frac{(2v')!}{v'!}
       \left(
         -\frac{r^{2}}{4}
       \right)^{v'}W_{n,v'}(a/r), &v=2v'.
     \end{array}
     \right.
  \end{align}
\end{subequations}
Setting $v=0,1,\ldots$ yields an infinite lower triangular system of
equations for $u_{q}(r)$:
\begin{align}
  \label{equ:system:1}
  \mathsf{E}\mathbf{U} = \mathbf{B},
\end{align}
with $\mathbf{U}=[u_{0}\,u_{1}\,\ldots]^{T}$ and the elements of
matrix $\mathsf{E}$ and vector $\mathbf{B}$ given by:
\begin{subequations}
  \label{equ:entries}
  \begin{align}
    E_{vq} &= 
    \left\{
    \begin{array}{ll}
      \J^{v}(q+1)v!E_{v-q,q}(r), & q\leq v;\\
      0, & q > v.
    \end{array}
    \right.\\
    B_{v} &= 
    \frac{\J}{4}\int_{0}^{1}T_{v}(r,a)s_{n}(a)\,\D a
  \end{align}
\end{subequations}
where
\begin{align}
  T_{v} &= 
  (-1)^{n+v'}
  v!
  \left(
    \frac{r}{2}
  \right)^{v}
  \sum_{m=0}^{\infty}
  A_{m}
  \left\{
    \begin{array}{lll}
      \displaystyle
      0 & v = 2v'+1, & v' < n+2m;\\
      \displaystyle
      \frac{V_{n+2m,v'-n-2m}(a/r)}{(v'-n-2m)!}
      & v = 2v'+1, & v' \geq n+2m;\\
      \displaystyle
      - \frac{\J}{v'!} 
      W_{n+2m,v'}(a/r) & v =
      2v'.
    \end{array}
  \right.
  \label{equ:tfunc}
\end{align}

Given a radial source term $s_{n}(a)$, Equation~\ref{equ:system:1} can
be solved to find the coefficients $u_{q}(r)$ of the equivalent line
source modes. Since it is lower triangular, the first few values of
$u_{q}$ can be reliably estimated, although ill-conditioning prevents
accurate solution for arbitrary large $q$.

\subsection{Radiated field}
\label{sec:radiated}

From the relationship between the radial source term and the line
source coefficients, some general properties of the acoustic field can
be stated. The first result, already shown in previous
work~\cite{carley10b,carley10c} is that, since the line source modes
with $q+1>k$ generate exponentially small fields, the acoustic field
has no more than $k$ degrees of freedom, in the sense that the
radiated field is given by a weighted sum of the fields due to no more
than $k$ elementary sources. From Equation~\ref{equ:system:1}, this
result can be extended.

The first extension comes from the fact that $B_{2v+1}\equiv0$, for
$v'<n$, on the right hand side of Equation~\ref{equ:system:1}. This
means that $u_{q}$, $q=2v'+1$, is uniquely defined by the lower order
coefficients with $q\leq 2v'$. The result is that the acoustic field
of azimuthal order $n$, whatever might be its radial structure, has no
more than $k-n$ degrees of freedom, whether in the near or far field.

A second extension comes from examination of
Equation~\ref{equ:system:1}. The first few entries of the system of
equations are:
\begin{align}
  \label{equ:system:a}
  \left[
    \begin{array}{rrrrr}
      1/2 & 0 & 0 & 0 & \cdots \\
      r/2 & 1/4 & 0 & 0 & \cdots \\
      \vdots &\vdots & \vdots & 0 & \cdots
    \end{array}
  \right]
  \left(
    \begin{array}{c}
      u_{0} \\ u_{1} \\ \vdots
    \end{array}
  \right)
  =
  \left(
    \begin{array}{c}
      B_{0} \\ 0 \\ \vdots
    \end{array}
  \right),
\end{align}
resulting in the solution:
\begin{align}
  \label{equ:system:sol}
  u_{0} = 2B_{0};\quad u_{1} = -2ru_{0} = -4rB_{0},
\end{align}
so that the ratio of $u_{0}$ and $u_{1}$ is constant, for arbitrary
$s_{n}(a)$. This means that low frequency sources of the same radius
and azimuthal order generate fields which vary only by a scaling
factor, since the higher order terms are exponentially small. Again,
this result holds in the near and in the far field.

Finally, if we attempt to isolate a source $s_{n}(a)$ associated with
a single line source mode, by setting $u_{q}\equiv1$ for some $q$,
with all other $u_{q}\equiv0$, we find that the line modes must occur
in pairs, since if $u_{2v'}\equiv1$, $u_{2v'+1}\neq0$, being fixed by
the condition $B_{2v'+1}\equiv0$, further reducing the number of
degrees of freedom or, alternatively, worsening the conditioning of
the inverse problem.

\subsection{Comparison to far-field methods}
\label{sec:far:field}

An alternative analysis which is widely used in radiation prediction
uses the far field approximations $R'\approx R - a\sin\phi\cos\psi$,
$1/R'\approx 1/R$. On this approximation:
\begin{align}
  \label{equ:far:field}
  P_{n} &\approx (-\J)^n\frac{\E^{\J k R}}{2R}
  \int_{0}^{1}
  J_{n}(ka\sin\phi) s_{n}(a) a\,\D a,
\end{align}
so that the radiated field is given by a Hankel transform of the
radial source, with a dependence on the polar angle $\phi$. In some
sense, this can also be viewed as fixing a limit on the radiated
information as in, for example, the use of ring sources to study
coherence effects on jet noise~\cite{michel09,michalke83}, or as a
spatial filter. The approach suffers, however, from its inability to
give information on the structure of the near field which might be of
use in understanding such experimental methods as near-field to
far-field correlations~\cite{laurendeau-jordan-delville-bonnet08}. The
approach presented in this paper gives the radiated field, near and
far, as the sum of products of two integrals. The first of these
integrals $\lfn{q}$ contains only radiation effects while the second
$u_{q}$ depends only on the source. The source and radiation terms are
thus `uncoupled', simplifying the problem of analysing the radiated
field, without needing to make a far-field approximation.

\section{Random disk source}
\label{sec:random}

The second problem considered is that of the noise radiated by a
random disk source. This is a general problem for broadband noise from
rotating systems and is also a model problem for jet noise, extending
the random ring source problem which has been studied previously in
order to examine the effects of source coherence on jet
noise~\cite{michalke83}. The assumptions made are that the source terms
are statistically stationary and that the statistical properties of
the source are symmetric about the source axis. It will also be
assumed that the non-dimensional wavenumber $k\lesssim2$, which is a
reasonable assumption for the frequency range of maximum noise level
for a subsonic jet. The result derived is an expression for the
cross-spectrum between the pressure at two points, which reduces to
the power spectrum when the points coincide. The expression is quite
general and, unlike previous formulae, does not require that the
points be in the acoustic far field of the source.

The starting point is an expression for the pressure radiated from a
source distributed over a unit disk:
\begin{align}
  \label{equ:disk:1}
  p(r,\theta,z,t) &= \int_{0}^{1}\int_{0}^{2\pi}
  \frac{q(a,\psi,t-R/c)}{4\pi R}a\,\D \psi\,\D a,
\end{align}
from which the correlation between $p$ measured at two points
$(r_{1},\theta_{1},z_{1})$ and $(r_{2},\theta_{2},z_{2})$ is:
\begin{align}
  \overline{p(r_{1},\theta_{1},z_{1},t)p(r_{2},\theta_{2},z_{2},t+\tau)}
  &=\nonumber\\
  \frac{1}{(4\pi)^{2}}
  \int_{0}^{1}\int_{0}^{2\pi}
  \int_{0}^{1}\int_{0}^{2\pi}
  \frac{\overline{q(a_{1},\psi_{1},t-R_{1}/c)q(a_{2},\psi_{2},t-R_{2}/c+\tau)}}
  {R_{1}R_{2}}a_{1}a_{2}
  \,\D \psi_{1}\,\D a_{1}
  \,\D \psi_{2}\,\D a_{2}.
\end{align}

Fourier transforming to find the cross spectrum between the points:
\begin{align}
  \label{equ:xspec}
  W_{12}(f) &= \frac{1}{(4\pi)^{2}}
  \int_{0}^{1}\int_{0}^{2\pi}
  \int_{0}^{1}\int_{0}^{2\pi}
  \frac{\E^{\J k(R_{2}-R_{1})}}{R_{1}R_{2}}
  Q_{12}(a_{1},\psi_{1};a_{2},\psi_{2}) a_{1}a_{2}
  \,\D \psi_{1}\,\D a_{1}
  \,\D \psi_{2}\,\D a_{2},\\
  Q_{12}(a_{1},\psi_{1};a_{2},\psi_{2}) &= \int_{-\infty}^{\infty} 
  \overline{q(a_{1},\psi_{1},t)q(a_{2},\psi_{2},t+\tau)}\E^{\J 2\pi f
    \tau}\,\D \tau,
\end{align}
where $Q_{12}$ is the correlation between the source at two points
$(a_{1},\psi_{1})$ and $(a_{2},\psi_{2})$, assumed real.

On the assumption of axial symmetry, the source correlation can depend
only on the angular separation between two points $\psi_{2}-\psi_{1}$,
so that $Q_{12}$ and $W_{12}$ can be expanded in Fourier series in
azimuth:
\begin{align*}
  Q_{12}(a_{1},\psi_{1};a_{2},\psi_{2}) &=
  \sum_{m=-\infty}^{\infty}Q_{12}^{(m)}(a_{1},a_{2})\E^{\J m
    (\psi_{2}-\psi_{1})},\\
  W_{12}(r_{1},\theta_{1},z_{1};r_{2},\theta_{2},z_{2}) 
  &=
  \sum_{m=-\infty}^{\infty}W_{12}^{(m)}(r_{1},z_{1};r_{2},z_{2})
  \E^{\J m (\theta_{2}-\theta_{1})},
\end{align*}
with:
\begin{align}
  \label{equ:xspec:2}
  W_{12}^{(m)}(r_{1},\theta_{1},z_{1};r_{2},\theta_{2},z_{2}) 
  &= \frac{1}{(4\pi)^{2}}
  \int_{0}^{1}\int_{0}^{2\pi}
  \frac{\E^{-\J (kR_{1}+m\psi_{1})}}{R_{1}}\\
  &\times\left[
    \int_{0}^{1}\int_{0}^{2\pi}
    \frac{\E^{\J (kR_{2}+m\psi_{2})}}{R_{2}}
    Q_{12}^{(m)}(a_{1},a_{2})
    a_{2}
    \,\D \psi_{2}\,\D a_{2}\,
  \right]
  a_{1}
  \,\D \psi_{1}\,\D a_{1}. \nonumber
\end{align}

Transforming to the equivalent line source form, as above:
\begin{align*}
  \frac{1}{4\pi}
  \int_{0}^{1}\int_{0}^{2\pi}
  \frac{\E^{\J (kR_{2}+m\psi_{2})}}{R_{2}}
  Q_{12}^{(m)}(a_{1},a_{2})
  a_{2}
  \,\D \psi_{2}\,\D a_{2}
  &=
  \sum_{q_{2}=0}^{\infty}
  u_{q_{2}}(r_{2},a_{1})\lfn{q_{2}}(k,r_{2},z_{2}),
\end{align*}
which results in:
\begin{align*}
  W_{12}^{(m)} &= \sum_{q_{2}} \lfn{q_{2}}(k,r_{2},z_{2})
  \frac{1}{4\pi}
  \int_{0}^{1}\int_{0}^{2\pi}
  \frac{\E^{-\J (kR_{1}+m\psi_{1})}}{R_{1}}
  u_{q_{2}}(r_{2},a_{1})
  a_{1}\,\D a_{1}\,\D\psi_{1},\\
  &= \sum_{q_{1}}\sum_{q_{2}} u_{q_{1}}(r_{1},r_{2})
  \lfn{q_{2}}(k,r_{2},z_{2})\lfn{q_{1}}^{*}(k,r_{1},z_{1}),
\end{align*}
where $*$ denotes complex conjugation. The coefficients $u_{q_{1}}$
are found by treating $u_{q_{2}}$ as the radial source in the
$(a_{1},\psi_{1})$ integral. Up to this point, the analysis is exact
but to simplify the development, we introduce the assumption $k<2$ so
that only modes of order~0 and~1 contribute to the acoustic field.

Solving Equation~\ref{equ:system:1} yields:
\begin{align*}
  u_{0} &= 2 B_{0},\\
  u_{1} &= 4(B_{1} - rB_{0}),
\end{align*}
with:
\begin{align*}
  B_{0} &= \int_{0}^{1}s_{m}(a)w_{m}(a/r)\,\D a,\\
  B_{1} &= \int_{0}^{1}s_{m}(a)v_{m}(a)\,\D a,
\end{align*}
where:
\begin{align*}
  w_{m}(x) &= \frac{1}{2\pi}\sum_{q=0}^{\infty}
  \frac{1}{q!} \frac{(2m+2q-1)!!(2q-1)!!}{2^{q}(2m+2q)!!}x^{m+2q+1},\\
  v_{m}(x) &= 
  \left\{
    \begin{array}{ll}
      x/2\pi, & m = 0;\\
      0, & m \neq 0.
    \end{array}
  \right.
\end{align*}

The result is that the $m$th azimuthal component of the cross-spectrum
between two field points for $k\lesssim2$ is given by:
\begin{align}
  W_{12}^{(m)} &= 
  \lfn{0}(k,r_{2},z_{2})
  \left[
    u_{00}\lfn{0}^{*}(k,r_{1},z_{1}) +
    u_{01}\lfn{1}^{*}(k,r_{1},z_{1})
  \right]
  \nonumber\\
  \label{equ:xspec:3}
  &+ \lfn{1}(k,r_{2},z_{2})
  \left[
    u_{10}\lfn{0}^{*}(k,r_{1},z_{1}) +
    u_{11}\lfn{1}^{*}(k,r_{1},z_{1})
  \right],
\end{align}
where
\begin{subequations}
  \label{equ:uij}
\begin{align}
  u_{00} &= 
  4\int_{0}^{1}\int_{0}^{1}
  Q_{12}^{(m)}(a_{1},a_{2})w_{m}(a_{1}/r_{1})w_{m}(a_{2}/r_{2})\,\D
  a_{1}\, \D a_{2},\\
  u_{01} &= 
  8\int_{0}^{1}\int_{0}^{1}
  Q_{12}^{(m)}(a_{1},a_{2})w_{m}(a_{2}/r_{2})
  [v_{m}(a_{1}) - r_{1}w_{m}(a_{1}/r_{1})]
  \,\D
  a_{1}\, \D a_{2},\\
  u_{10} &= 
  8\int_{0}^{1}\int_{0}^{1}
  Q_{12}^{(m)}(a_{1},a_{2})w_{m}(a_{1}/r_{1})
  [v_{m}(a_{2}) - r_{2}w_{m}(a_{2}/r_{2})]
  \,\D
  a_{1}\, \D a_{2},\\
  u_{11} &= 
  16\int_{0}^{1}\int_{0}^{1}
  Q_{12}^{(m)}(a_{1},a_{2})
  [v_{m}(a_{1}) - r_{1}w_{m}(a_{1}/r_{1})]
  [v_{m}(a_{2}) - r_{2}w_{m}(a_{2}/r_{2})]
  \,\D
  a_{1}\, \D a_{2}.
\end{align}  
\end{subequations}
The modal coefficients of the cross-spectrum of a jet noise field, at
the wavenumbers of interest in practice, are thus fixed by four
coefficients, functions of the radial separations $r_{1}$ and $r_{2}$,
which are weighted integrals of the source cross-spectrum.

\section{Results}
\label{sec:results}


To check the analyses presented above, and to consider their
implications, some results are presented for tonal and random disk
sources.

\subsection{Line source coefficient evaluation}
\label{sec:coefficient}

\begin{figure}
  \centering
  \begin{tabular}{c}
    \includegraphics{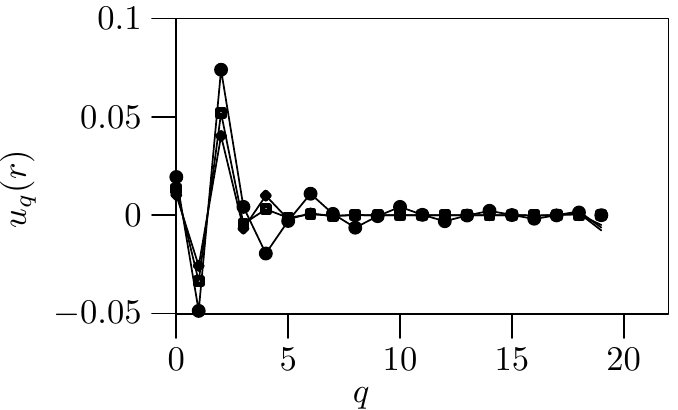} \\
    $n=2$ \\
    \includegraphics{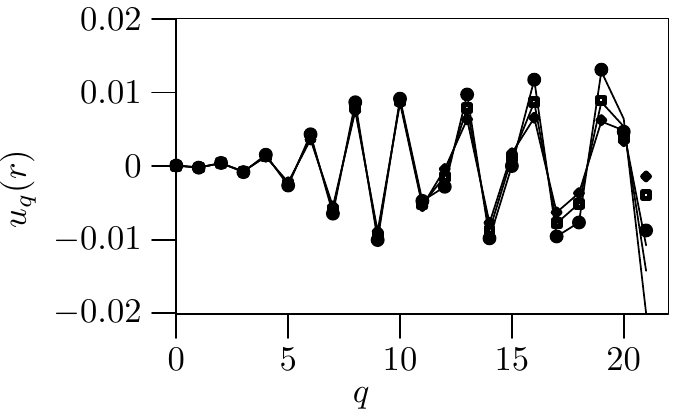} \\
    $n=16$
  \end{tabular}
  \caption{Line source mode coefficients computed using the method of
    section~\ref{sec:coefficients} (solid lines) and directly from
    analytical formulae (symbols) for $r=5/4$, $s=r_{1}^{\gamma}$,
    $\gamma=0$ (circles), $\gamma=2$ (squares) and $\gamma=4$
    (diamonds) for $n=2$ and $16$.}
  \label{fig:cfft:compare}
\end{figure}

The first results are a check on the calculation of the coefficients
$u_{q}(r)$ comparing those computed using Equation~\ref{equ:system:1}
and those computed directly from exact closed-form
expressions~\cite{carley99} for $K(r,r_{2})$ in the case when the
radial source term is a monomial in radius
$s_{n}=r_{1}^{\gamma}$. Figure~\ref{fig:cfft:compare} compares the two
sets of coefficients for $\gamma=0,2,4$, with the plots terminated at
a value of $q$ where the difference between the two sets of results
becomes noticeable, $q\approx20$. This gives an indication of the
effect of the ill conditioning of Equation~\ref{equ:system:1}. For
$q\lesssim 20$, the computed values of $u_{q}$ are reliable. It is
noteworthy that for small $q$, the coefficients are practically equal
for all values of $\gamma$ so that for low frequency radiation, the
radiated fields will be practically indistinguishable.

\subsection{Tonal radiation from a disk}
\label{sec:disk}

\begin{figure*}
  \centering
  \begin{tabular}{cc}
    \includegraphics{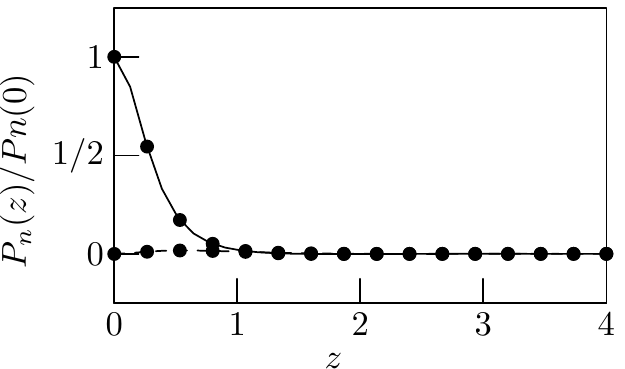} &
    \includegraphics{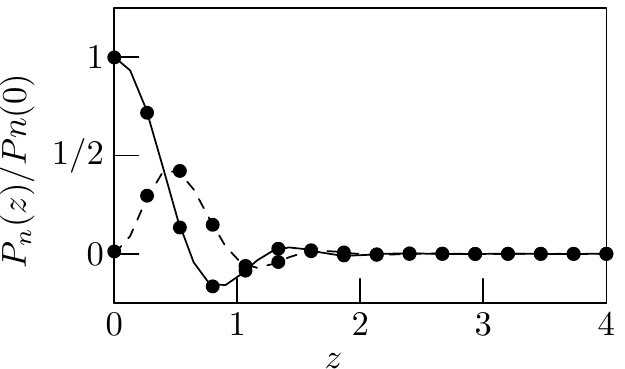} 
  \end{tabular}
  \caption{Acoustic field predicted by full numerical integration
    (lines) and line source summation (symbols) for $n=8$,
    $r=5/4$. Real part shown solid, imaginary part dashed. Left hand
    plot: $k=5$; right hand plot: $k=9$.}
  \label{fig:disk}
\end{figure*}

As a test of the ability to predict radiation from tonal sources, we
present data for the acoustic field of a disk source with $n=8$,
$s_{n}=J_{n}(a_{n1}a)$, where $a_{n1}$ is the first non-zero root of
$J_{n}(x)$. Full numerical integration and line source calculations
have been performed for two wavenumbers, $k=5$ and $k=9$,
respectively. The first~11 line source modes were used in each case,
with the modal coefficients being found from
Equation~\ref{equ:system:1}. Sample results are shown in
Figure~\ref{fig:disk}, with the data scaled on the value at $z=0$, and
it is clear that the line source model gives accurate results, even
when only a subset of the modes is used. From these, and other, data,
the reliability of the model for tonal sources is confirmed.



\subsection{Low frequency random source}
\label{sec:low:random}

In order to generate data to test the random disk source model, we
must assume a form for the source correlation.
Michalke~\cite{michalke83} gives a form suitable for a ring source
which meets the symmetry requirements laid out above. With the
addition of radially varying terms, Michalke's expression can be
extended:
\begin{align}
  \label{equ:coherence:1}
  Q_{12}(a_{1},\psi_{1};a_{2},\psi_{2}) &= 
  q(a_{1})q(a_{2})
  \exp
  \left[
    -\frac{(a_{1}-a_{2})^{2}}{\beta^{2}}
  \right]
  \exp 
  \left[
    -\frac{1-\cos(\psi_{1}-\psi_{2})}{\alpha^{2}}
  \right]
\end{align}
with $\alpha$ being an azimuthal length scale $\beta$ controlling the
correlation in radius. Equation~\ref{equ:coherence:1} can be
interpreted as the product of the local source strengths $q(a_{1})$
and $q(a_{2})$ with a coherence function, given by the exponentials,
which is symmetric in source position and has unit value when the
source points coincide.

The azimuthal components of $Q_{12}$ can be found from mathematical
tables~\cite{gradshteyn-ryzhik80,michalke83} as:
\begin{align}
  \label{equ:coherence:2}
  Q_{12}^{(m)} &= q(a_{1})q(a_{2})
  \exp
  \left[
    -\frac{(a_{1}-a_{2})^{2}}{\beta^{2}}
  \right]
  \exp
   \left[
     -\frac{1}{\alpha^{2}}
   \right]
  I_{m}(1/\alpha^{2})
\end{align}
where $I_{m}$ is a modified Bessel function.

\begin{figure*}
  \centering
  \begin{tabular}{cc}    
    \includegraphics{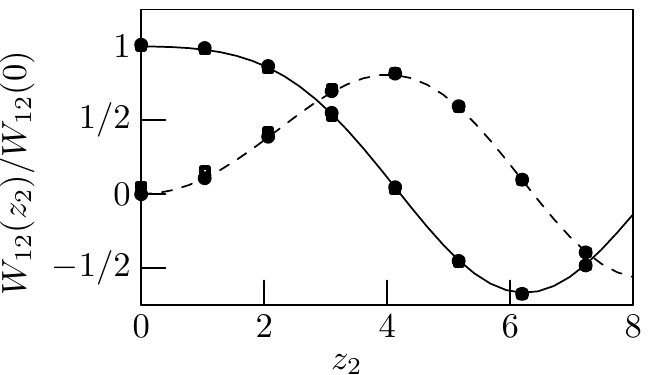} &
    \includegraphics{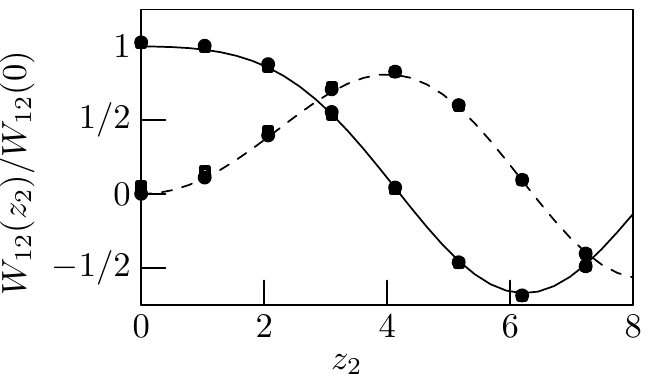} \\
    \textit{a} & \textit{b} \\
    \includegraphics{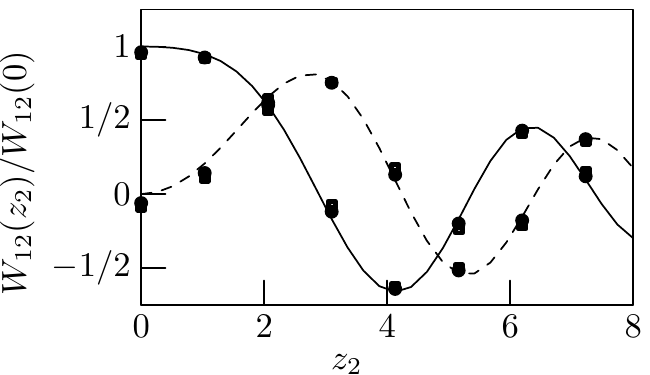} &
    \includegraphics{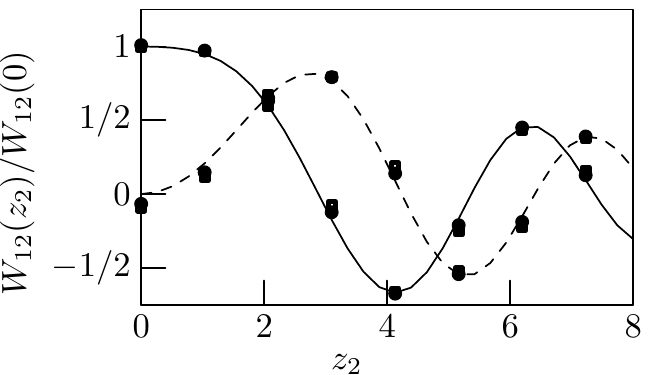} \\
    \textit{c} & \textit{d}
  \end{tabular}
  \caption{Cross-spectrum $W_{12}^{(m)}(z_{2})$ scaled on
    $W_{12}^{(m)}(0)$, $r_{1}=5/4$, $z_{1}=0$, $r_{2}=5$. Numerical
    evaluation shown as solid line (real part) and dashed line
    (imaginary part); Equation~\ref{equ:xspec:3} with numerical
    evaluation of $L_{q}$ shown as circles; Equation~\ref{equ:xspec:3}
    with far-field approximation shown as squares. Parameters:
    \textit{a}: $k=1$, $m=0$, $\alpha=1$, $\beta=100$;
    \textit{b}: $k=1$, $m=0$, $\alpha=3$, $\beta=0.01$;
    \textit{c}: $k=2$, $m=1$, $\alpha=1$, $\beta=100$;
    \textit{d}: $k=2$, $m=1$, $\alpha=3$, $\beta=0.01$.}
  \label{fig:xspec}
\end{figure*}

Figure~\ref{fig:xspec} shows sample results for the predicted cross
spectrum between pressure at a point $r_{1}=5/4$, $z_{1}=0$ and
$r_{2}=5$, $0\leq z_{2}\leq8$, for a disk source of unit strength. The
reference results are the cross-spectra found by full numerical
integration of Equation~\ref{equ:xspec:2}. The first comparison is
with Equation~\ref{equ:xspec:3} where the functions $L_{q}$ have been
evaluated by numerical integration. In the second comparison, the
functions $L_{q}$ have been evaluated using the exact in-plane result,
Equation~\ref{equ:line:series}, for $z_{1}=0$, and the far-field
approximation, Equation~\ref{equ:lfunc:ff}, for $r_{2}=5$, $0\leq
z_{2}\leq 8$. All data have been scaled on the numerically evaluated
cross-spectrum at $z_{2}=0$.

The first obvious point from Figure~\ref{fig:xspec} is the similarity
of the cross-spectra, even for quite large variations in the parameter
$\beta$: changing $m$ changes the form of the radiated field, as might
be expected, but changes in the source correlation have little effect
on the radiated field. The second point is that the line source
approach gives very good results, even for $k=2$ where, in principle,
the approximation used should start to break down. Finally, although
computational efficiency is not the primary aim of the method, we note
that the line source approach converts the four dimensional integral,
Equation~\ref{equ:xspec:2}, required at each field point, into four
two-dimensional integrals which are functions of radial separation
only, Equation~\ref{equ:uij}, and four one-dimensional integrals
$\lfn{i}$, giving a large saving in calculation time.

\subsection{Noise cancellation by an equivalent source}
\label{sec:cancellation}

One implication of the results of this paper is that it is not
possible to tell different sources apart if, to within a scaling
factor, they have same line source coefficients $u_{q}$, for those
line source modes with $q<k$. Even without considering errors from
background noise or other causes, this is equivalent to a condition on
weighted integrals of the radial source $s_{n}$. Any sources which
yield the same, or nearly the same, integrals $B_{v}$ for $v<V$, with
$V$ a positive integer, in Equation~\ref{equ:entries}, will have
indistinguishable acoustic fields for $k<V$.

This conclusion can also be read as a statement about noise
cancellation, such as in active noise control. The acoustic field of a
given source can be cancelled by any source which has the same set of
line source coefficients. 

\begin{figure}
  \centering
    \includegraphics{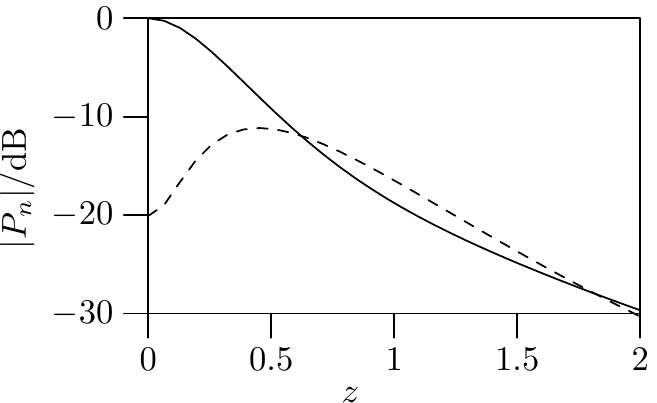} \\
    \includegraphics{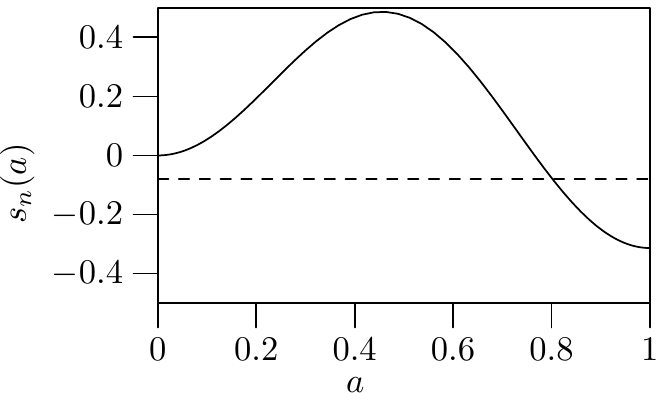} 
    \caption{Cancellation effects for radial source terms with $n=2$,
      $k=1$, $r=5/4$: top figure radiated field from original
      $s_{n}(a)$ (solid) and modified source $s_{n}(a)-\zeta
      s_{n}'(a)$ (dashed); bottom row source terms $s_{n}(a)$ (solid)
      and $\zeta s_{n}'(a)$ dashed.}
  \label{fig:field:compare}
\end{figure}

An example of this cancellation is shown in
Figure~\ref{fig:field:compare}. The original field is generated using
a source term $s_{n}(a)$ and the line source coefficients $u_{q}$ of
$s_{n}$ are calculated. A secondary source term $s_{n}'(a)$ is
generated and its line source coefficients $u_{q}'$ are computed. The
secondary source $s_{n}'$ is then scaled by a factor
$\zeta=u_{0}/u_{0}'$. As a test, $s_{n}=J_{n}(a_{n2}a)$, with $a_{n2}$
the second extremum of $J_{n}(x)$, and $s_{n}'\equiv1$. The first plot
in Figure~\ref{fig:field:compare} shows the field due to $s_{n}$ and
that radiated by $s_{n}-\zeta s_{n}'$. The large reduction,
20\deci\bel, in the radiation near the source plane is obvious,
although there is a small increase in the noise field around
$z=1$. The source terms are shown in the second plot of
Figure~\ref{fig:field:compare}. The secondary source $\zeta s_{n}'$ is
of much smaller amplitude than $s_{n}$ even though it generates a
nearly-equivalent field: the effect of matching the line source
coefficients has been to produce a field which is very similar to that
of the original source, even though the source distributions are quite
different in form and in amplitude.

\section{Conclusions}
\label{sec:conclusions}

The radiation properties of disk sources of arbitrary radial variation
have been analyzed to establish the part of the source which radiates
into the acoustic field, without recourse to a far field
approximation. Limits have been established on the number of degrees
of freedom of the part of the source which radiates and the
implications of these limits have been discussed for the problems of
rotor noise and studies of source mechanisms in jets. The analysis has
been developed for tonal and for random sources, with implications for
applications in active control of noise from rotors and experimental
analysis of jet noise sources. 


\begin{thebibliography}{10}

\bibitem{lighthill52}
M.~J. Lighthill.
\newblock On sound generated aerodynamically: {I General} theory.
\newblock {\em Proceedings of the Royal Society of London. \textrm{A}.},
  211:564--587, 1952.

\bibitem{ffowcs-williams-hawkings69b}
J.~E. {Ffowcs Williams} and D.~L. Hawkings.
\newblock Sound generation by turbulence and surfaces in arbitrary motion.
\newblock {\em Philosophical Transactions of the Royal Society of London.
  \textrm{A}.}, 264:321--342, 1969.

\bibitem{oberhettinger61a}
F.~Oberhettinger.
\newblock On transient solutions of the ``baffled piston'' problem.
\newblock {\em Journal of Research of the National Bureau of Standards B},
  65(1):1--6, January--March 1961.

\bibitem{pierce89}
A.~D. Pierce.
\newblock {\em Acoustics: {An} introduction to its physical principles and
  applications}.
\newblock Acoustical Society of America, New York, 1989.

\bibitem{mellow06}
Tim Mellow.
\newblock On the sound field of a resilient disk in an infinite baffle.
\newblock {\em Journal of the Acoustical Society of America}, 120(1):90--101,
  2006.

\bibitem{mellow08}
Tim Mellow.
\newblock On the sound field of a resilient disk in free space.
\newblock {\em Journal of the Acoustical Society of America},
  123(4):1880--1891, 2008.

\bibitem{gutin48}
L.~Gutin.
\newblock On the sound field of a rotating propeller.
\newblock Technical Memorandum 1195, NACA, Langley Aeronautical Laboratory,
  Langley Field, Va. USA, 1948.

\bibitem{wright69}
S.~E. Wright.
\newblock Sound radiation from a lifting rotor generated by asymmetric disk
  loading.
\newblock {\em Journal of Sound and Vibration}, 9(2):223--240, 1969.

\bibitem{chapman93}
C.~J. Chapman.
\newblock The structure of rotating sound fields.
\newblock {\em Proceedings of the Royal Society of London. \textrm{A}.},
  440:257--271, 1993.

\bibitem{carley99}
M.~Carley.
\newblock Sound radiation from propellers in forward flight.
\newblock {\em Journal of Sound and Vibration}, 225(2):353--374, 1999.

\bibitem{gerard-berry-masson05a}
Anthony G\'{e}rard, Alain Berry, and Patrice Masson.
\newblock Control of tonal noise from subsonic axial fan. {Part} 1:
  reconstruction of aeroacoustic sources from far-field sound pressure.
\newblock {\em Journal of Sound and Vibration}, 288:1049--1075, 2005.

\bibitem{gerard-berry-masson05b}
Anthony G\'{e}rard, Alain Berry, and Patrice Masson.
\newblock Control of tonal noise from subsonic axial fan. {Part} 2: active
  control simulations and experiments in free field.
\newblock {\em Journal of Sound and Vibration}, 288:1077--1104, 2005.

\bibitem{gerard-berry-masson-gervais07}
A.~G\'{e}rard, A.~Berry, P.~Masson, and Y.~Gervais.
\newblock Evaluation of tonal aeroacoustic sources in subsonic fans using
  inverse models.
\newblock {\em AIAA Journal}, 45(1):98--109, 2007.

\bibitem{peake-boyd93}
N.~Peake and W.~K. Boyd.
\newblock Approximate method for the prediction of propeller noise near-field
  effects.
\newblock {\em Journal of Aircraft}, 30(5):603--610, 1993.

\bibitem{li-zhou95}
{Xiao-dong} Li and Sheng Zhou.
\newblock Spatial transformation of discrete sound field radiated by
  propellers.
\newblock In {\em First Joint CEAS/AIAA Aeroacoustics Conference}, pages
  1241--1250, 1995.

\bibitem{minniti-blake-mueller01a}
R.~J. Minniti, W.~K. Blake, and T.~J. Mueller.
\newblock Inferring propeller inflow and radiation from near-field response,
  part 1: {Analytic} development.
\newblock {\em AIAA Journal}, 39(6):1030--1036, June 2001.

\bibitem{minniti-blake-mueller01b}
R.~J. Minniti, W.~K. Blake, and T.~J. Mueller.
\newblock Inferring propeller inflow and radiation from near-field response,
  part 2: {Empirical} application.
\newblock {\em AIAA Journal}, 39(6):1037--1046, June 2001.

\bibitem{holste-neise97}
F.~Holste and W.~Neise.
\newblock Noise source identification in a propfan model by means of acoustical
  near field measurements.
\newblock {\em Journal of Sound and Vibration}, 203(4):641--665, 1997.

\bibitem{lewy05}
Serge Lewy.
\newblock Inverse method predicting spinning modes radiated by a ducted fan
  from free-field measurements.
\newblock {\em Journal of the Acoustical Society of America}, 117(2):744--750,
  2005.

\bibitem{lewy08}
Serge Lewy.
\newblock Numerical inverse method predicting acoustic spinning modes radiated
  by a ducted fan from free-field test data.
\newblock {\em Journal of the Acoustical Society of America}, 124(1):247--256,
  2008.

\bibitem{castres-joseph07a}
Fabrice~O. Castres and Philip~F. Joseph.
\newblock Mode detection in turbofan inlets from near field sensor arrays.
\newblock {\em Journal of the Acoustical Society of America}, 121(2):796--807,
  February 2007.

\bibitem{castres-joseph07b}
Fabrice~O. Castres and Philip~F. Joseph.
\newblock Experimental investigation of an inversion technique for the
  determination of broadband duct mode amplitudes by the use of near-field
  sensor arrays.
\newblock {\em Journal of the Acoustical Society of America}, 122(2):848--859,
  2007.

\bibitem{carley09}
Michael Carley.
\newblock Inversion of spinning sound fields.
\newblock {\em Journal of the Acoustical Society of America}, 125(2):690--697,
  2009.

\bibitem{carley10b}
Michael Carley.
\newblock Analysis of the radiated information in spinning sound fields.
\newblock {\em Journal of the Acoustical Society of America}, 2010, in press.

\bibitem{carley10c}
Michael Carley.
\newblock The near-field of spinning sources: Why source identification is
  hard.
\newblock {\em Proceedings of Meetings on Acoustics}, 9(1):040004--040004--11,
  2010.

\bibitem{freund01}
Jonathan~B. Freund.
\newblock Noise sources in a low-{Reynolds}-number turbulent jet at {Mach} 0.9.
\newblock {\em Journal of Fluid Mechanics}, 438:277--305, 2001.

\bibitem{freund10}
Jonathan~B. Freund.
\newblock Adjoint-based optimization for understanding and suppressing jet
  noise.
\newblock In {\em IUTAM Symposium on Computational Aero-Acoustics (CAA) for
  Aircraft Noise Prediction}, 2010.
\newblock \url{http://www.southampton.ac.uk/~gabard/IUTAM/}, (last viewed 23
  July 2010).

\bibitem{jordan-schlegel-stalnov-noack-tinney07}
P.~Jordan, M.~Schlegel, O.~Stalnov, B.~R. Noack, and C.~E. Tinney.
\newblock Identifying noisy and quiet modes in a jet.
\newblock In {\em 13th AIAA/CEAS Aeroacoustics Conference}, 2007.
\newblock AIAA 2007-3602.

\bibitem{michel07}
Ulf Michel.
\newblock Influence of source interference on the directivity of jet mixing
  noise.
\newblock In {\em 13th AIAA/CEAS Aeroacoustics Conference}, 2007.
\newblock {AIAA}-2007-3648.

\bibitem{michel09}
Ulf Michel.
\newblock The role of source interference in jet noise.
\newblock In {\em 15th AIAA/CEAS Aeroacoustics Conference}, 2009.
\newblock {AIAA} 2009-3377.

\bibitem{sinayoko-agarwal10}
Samuel Sinayoko and Anurag Agarwal.
\newblock Flow filtering and the physical sources of aerodynamic sound.
\newblock In {\em IUTAM Symposium on Computational Aero-Acoustics (CAA) for
  Aircraft Noise Prediction}, 2010.
\newblock \url{http://www.southampton.ac.uk/~gabard/IUTAM/}, (last viewed 23
  July 2010).

\bibitem{stepanishen76}
Peter~R. Stephanishen.
\newblock Asymptotic behavior of the acoustic nearfield of a circular piston.
\newblock {\em Journal of the Acoustical Society of America}, 59(4):749--754,
  1976.

\bibitem{michalke83}
A.~Michalke.
\newblock Some remarks on source coherence affecting jet noise.
\newblock {\em Journal of Sound and Vibration}, 87(1):1--17, 1983.

\bibitem{carley10}
Michael Carley.
\newblock Series expansion for the sound field of a ring source.
\newblock {\em Journal of the Acoustical Society of America}, 2010, in press.

\bibitem{goldstein74}
M.~Goldstein.
\newblock Unified approach to aerodynamic sound generation in the presence of
  solid boundaries.
\newblock {\em Journal of the Acoustical Society of America}, 56(2):497--509,
  1974.

\bibitem{gradshteyn-ryzhik80}
I.~Gradshteyn and I.~M. Ryzhik.
\newblock {\em Table of integrals, series and products}.
\newblock Academic, London, 5th edition, 1980.

\bibitem{laurendeau-jordan-delville-bonnet08}
E.~Laurendeau, P.~Jordan, J.~Delville, and J.-P. Bonnet.
\newblock Source-mechanism identification by nearfield-farfield pressure
  correlations in subsonic jets.
\newblock {\em International Journal of Aeroacoustics}, 7(1):41--68, 2008.

\end{thebibliography}

\end{document}